\documentclass[printer]{aa}  
\usepackage{graphicx}
\usepackage{txfonts}
\usepackage{verbatim}
 \usepackage{longtable,lscape}
 \usepackage{epsfig}
\usepackage{natbib}
\usepackage{xspace}

\usepackage{setspace}\usepackage{threeparttable}

\bibpunct{(}{)}{;}{a}{}{,} 

\newcommand{\nodata}{~$\cdots$~}

\newcommand{\eprest}{$E_{\rm p,rest}$\xspace}
\newcommand{\ep}{$E_{\rm p}$\xspace}
\newcommand{\eiso}{$E_{\rm{iso}}$\xspace}
\newcommand{\tninety}{$T_{90}$\xspace}
\newcommand{\tninetyrest}{$T_{90, \rm{rest}}$\xspace}

\newcommand{\batse}{{BATSE}\xspace}

\begin{document}

\title{Rest-frame properties of 32 gamma-ray bursts \\
       observed by the \emph{Fermi} Gamma-Ray Burst Monitor}


   \author{  D.Gruber\inst{1},
		J. Greiner\inst{1},
		A. von Kienlin\inst{1},
		A. Rau\inst{1},
		M. S. Briggs\inst{2},
		V. Connaughton\inst{2},
		A. Goldstein\inst{2},
		A. J. van der Horst\inst{2},
		M. Nardini\inst{1},							
		P. N. Bhat\inst{2},
		E. Bissaldi\inst{1},		
		J. M. Burgess\inst{2},
		V. L. Chaplin\inst{2},		
		R. Diehl\inst{1},
		G. J. Fishman\inst{4},
		G. Fitzpatrick\inst{3}, \\
		S. Foley\inst{1},
		M. H. Gibby\inst{5},
		M. M. Giles\inst{5},		
		S. Guiriec\inst{2},
		R. M. Kippen\inst{6},
		C. Kouveliotou\inst{4},
		L. Lin\inst{2},
		S. McBreen\inst{1,3},\\
		C. A. Meegan\inst{7},
		F. Olivares E.\inst{1},
		W. S. Paciesas\inst{2},
		R. D. Preece\inst{2},
		D. Tierney\inst{3},
		C. Wilson-Hodge\inst{4}.
          }

   \institute{Max Planck Institute for extraterrestrial Physics, 
   		Giessenbachstrasse 1, 85748, Garching, Germany\\
		\email{dgruber@mpe.mpg.de}
		\and
		University of Alabama in Huntsville, NSSTC, 
		320 Sparkman Drive, Huntsville, AL 35805, USA 
		\and
		University College, Dublin, Belfield, Stillorgan Road, Dublin 4, Ireland
		\and
		Space Science Office, VP62, NASA/Marshall Space Flight Center
		Huntsville, AL 35812, USA
		\and
		Jacobs Technology, Inc., Huntsville, Alabama
		\and
		 Los Alamos National Laboratory, 
		 P.O. Box 1663, Los Alamos, NM 87545, USA
		 \and
		 Universities Space Research Association, 
		 NSSTC, 320 Sparkman Drive, Huntsville, AL 35805, USA
		  }


 
  \abstract
  {}
  {In this paper we study the main spectral and temporal properties of gamma-ray bursts (GRBs) observed by \emph{Fermi}/GBM. We investigate these key properties of GRBs in the rest-frame of the progenitor and test for possible intra-parameter correlations to better understand the intrinsic nature of these events.}
  {Our sample comprises 32 GRBs with measured redshift that were observed by GBM until August 2010. 28 of them belong to the long-duration population and 4 events were classified as short/hard bursts. For all of these events we derive, where possible, the intrinsic peak energy in the $\nu F_{\nu}$ spectrum (\eprest), the duration in the rest-frame, defined as the time in which 90\% of the burst fluence was observed (\tninetyrest) and the isotropic equivalent bolometric energy  (\eiso).}
  {The distribution of \eprest has mean and median values of $1.1$~MeV and $750$~keV, respectively. A log-normal fit to the sample of long bursts peaks at $\sim 800$~keV. No high-\ep population is found but the distribution is biased against low \ep values. We find the lowest possible \ep that GBM can recover to be $\approx 15$~keV. The \tninetyrest distribution of long GRBs peaks at $\sim 10$~s. The distribution of \eiso has mean and median values of $8.9\times 10^{52}$~erg and $8.2 \times 10^{52}$~erg, respectively. We confirm the tight correlation between \eprest and \eiso (Amati relation) and the one between \eprest and the 1-s peak luminosity ($L_p$) (Yonetoku relation). Additionally, we observe a parameter reconstruction effect, i.e. the low-energy power law index $\alpha$ gets softer when \ep is located at the lower end of the detector energy range. Moreover, we do not find any significant cosmic evolution of neither \eprest nor \tninetyrest.}
  {}


  \titlerunning{Rest-frame properties of 32 gamma-ray bursts observed by the \emph{Fermi} Gamma-Ray Burst Monitor}
  \authorrunning{D. Gruber et al.}

  \maketitle
%

\section{Introduction}

Gamma-ray Bursts (GRB) are the most luminous flashes of $\gamma$-rays known to humankind. It is generally believed that they originate from a compact source with highly relativistic collimated outflows ($\Gamma > 100$).

A large fraction of our knowledge of the prompt emission comes from the Burst and Transient Explorer \citep[\batse,][]{meegan92} onboard the Compton Gamma-Ray Observatory (\emph{CGRO}, 1991-2000). Unfortunately, only a handful of \batse bursts had a measured redshift because the \batse error boxes were too large and thus, follow-up observations with X-ray and optical instrumentation was very limited. Moreover, the first afterglow was only detected in 1997, already near the end of the \batse mission. The lack of distance measurements led to a focus of GRB studies in the observer frame without redshift corrections. Due to the cosmological origin of GRBs, such a correction
is likely to be necessary to understand the intrinsic nature of these events.

With the two dedicated satellites, Beppo-\emph{SAX} \citep{boella97} and \emph{Swift} \citep{gehrels04}, the situation has changed and afterglow and host galaxy spectroscopy has provided redshifts for more than 200 events by now. Unfortunately, the relatively narrow energy band of Beppo-\emph{SAX} (0.1~keV - 300~keV) and \emph{Swift}/BAT (15~keV - 150~keV) limits the constraints on the prompt emission spectrum because the peak energy, \ep, in the $\nu F_{\nu}$ spectrum of GRBs, can only be determined for low \ep values and is often unconstrained \citep[][]{butler07,saka09}.

In this work we will take advantage of the broad energy coverage of the \emph{Fermi}/GBM (8~keV - 40~MeV) to study the primary spectral and temporal properties, such as \ep, \tninety, the time interval in which 90\% of the burst fluence has been observed, and \eiso, the isotropic equivalent bolometric energy, in the rest-frame of the progenitors of 32 GRBs with measured redshift.

\section{GRB sample and analysis}

\subsection{Instrumentation}

The Gamma-Ray Burst Monitor (GBM) is one of the instruments onboard the
\emph{Fermi} Gamma-Ray Space Telescope \citep{atwood09} launched on
June 11, 2008. Specifically designed for GRB studies, GBM observes the whole unocculted sky with a total of 12 thallium-activated sodium iodide (NaI(Tl)) scintillation detectors
covering the energy range from 8 keV to 1~MeV and
two bismuth germanate scintillation detectors (BGO) sensitive to energies between
150~keV and 40~MeV \citep{meegan09}.
Thus, GBM offers a unprecedented view of GRBs, covering more then 3 decades in energy.


\subsection{Burst selection}

The selection criterion for our sample is solely based on the
redshift determination. We form a sample of 32 bursts detected by GBM up to October 16th, 2010, with
known redshift (determined either spectroscopically or photometrically).

\begin{figure}[htbp]
\centering%
\includegraphics[width=0.5\textwidth, clip]{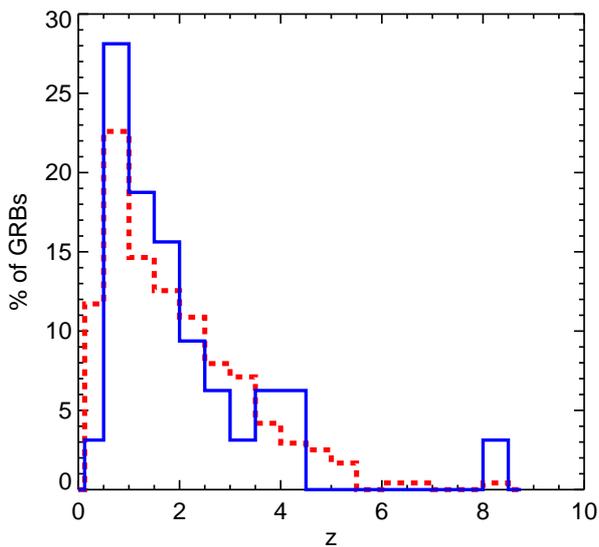}%
\caption{Redshift distribution in \% of GBM GRBs (blue solid line) compared to all 239 GRBs with measured redshift to date (red dashed line). Both samples contain long and short bursts.}
\label{fig:histoz}
\end{figure}

Our sample contains 4 short and 28 long GRBs.
The redshift distribution of the GBM GRBs is shown in Fig.~\ref{fig:histoz} together
with a histogram of all 239 GRBs with redshift determinations to date\footnote{\tt{www.mpe.mpg.de/\textasciitilde jcg/grbgen.html}}. A two-tailed Kolmogorov-Smirnov (KS) test between the full sample and the GBM-only sample shows that the two distributions are very similar ($P=84$~\%). In conclusion, the GBM-only sample is representative of the full GRB sample with redshift.

\subsection{Analysis of the GBM Data}

For the \ep determination, CSPEC data \citep{meegan09} with a time resolution of 1.024~s (4.096~s pre-trigger)
were used. For the short GRBs, i.e. those with $T_{90} \le 2$~s, time-tagged event  \citep[TTE,][]{meegan09} data
were used with a fine time resolution of 64~ms and 
the same channel boundaries as CSPEC data. 

Detectors with source angles
greater than 60$^\circ$ and those occulted by the spacecraft or solar
panels were discarded. A maximum of three NaI detectors were used for
each $E_p$ determination. In four cases (GRB~090929A, GRB~090904B,
GRB~090618, and GRB~081007) only one NaI detector could be used due to
the reasons mentioned above.
Where possible, both BGO detectors were included in the analysis if
they were not occulted by the satellite during the prompt
emission. Even if there was no apparent signal in the BGOs, they can
help to determine an upper limit of the GRB signal in the spectra.

The spectral analysis was performed with the software 
package RMFIT\footnote{RMFIT 
for GBM and LAT analysis was developed by the GBM Team and is publicly available at
{\tt {fermi.gsfc.nasa.gov/ssc/data/analysis/}}.} (version 3.3rc8) and the GBM Response Matrices v1.8. 
To account for the changing orientation of the source with respect to the detectors caused by the slew
of the spacecraft, the detector response matrices (DRM) were generated for every 2 degrees on the sky.

For each burst 
we fitted for every energy channel a low-order polynomial to a user defined background interval 
before and after the prompt emission 
and interpolated this fit across the source interval.

Three model fits
were applied to all time bins with a signal-to-noise (S/N) of at least 3.5 above
the background model: a single power-law (PL), a power law function with an
exponential high energy cutoff (COMP) and the Band function
\citep{band93}. For three GRBs (GRB~090424, GRB~090618, and
GRB~090926A) an effective area correction was applied to the BGO with
respect to the NaI detectors to account for systematics which dominate the statistical errors due to
the brightness of these events. The best model fit is the function which provides
the best Castor C-stat value \citep{cash79}. An improvement by $\Delta$C-stat$=10$ for every degree of freedom is required. The profile of the Cash
statistics was used to estimate the 1$\sigma$ asymmetric error.

The values obtained with this analysis method may be superseded by the GBM spectral catalogue released by the GBM team (Goldstein et al., in preparation).

\section{The intrinsic properties}
\subsection{The peak energy (\ep)}
\label{subsec:ep}

\subsubsection{Instrumental bias}
\label{subsubsec:instbias}

\begin{figure}[htbp]
\begin{center}
\includegraphics[width=0.5\textwidth]{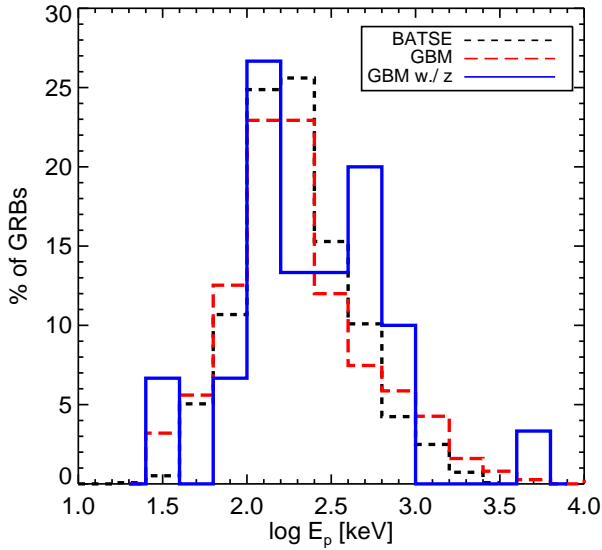}
\caption{ \ep values of 1367 \batse/CGRO GRBs with a log-normal fit (black dotted histogram and line), 375 GBM GRBs and log-normal fit (red dashed histogram and line) and 30 GBM GRBs with redshift measurement (blue solid histogram). A KS test suggests that all three samples were drawn from the same distribution.}
\label{fig:epbatse}
\end{center}
\end{figure}

A first important issue which needs to be addressed is whether the bursts observed by GBM are drawn from the same distribution as the bursts which were observed by \batse. Because of the broader sensitivity of GBM to higher energies, there could be a significant deviation towards higher \ep values. To answer this question, we use the \batse catalogue\footnote{\tt{www.batse.msfc.nasa.gov/~goldstein/}} from which we extract the \ep values which were obtained from the time averaged spectra (fluence spectral fits). Ignoring the power-law (PL), Gaussian Log and Smoothly Broken Power Law (SBPL) fits, we elected to the COMP model or Band function for the purpose of this test. We require $\Delta E_{\rm{p}} /{E_{\rm{p}}} \le 0.4$ and the low-energy power-law index, $\alpha$ has an absolute error $\sigma_{\alpha} \le 0.4$. GRBs which did not fulfill these criteria were rejected from the sample.  The Band function was always preferred over the COMP model in cases in which the high-energy power-law index $\beta$ was constrained ($\sigma_{\beta} < 0.4$). Both long and short GRBs were included in this analysis.
The so obtained \ep distribution shown in Fig. \ref{fig:epbatse}.

The same selection cut was applied to the 2-year GBM spectral catalogue. The red histogram in Fig. \ref{fig:epbatse} shows this distribution. Both distributions peak at $\approx$~170 keV and show the same standard deviation. A KS test reveals that the difference between the two samples is not statistically meaningful ($P=18$\%). This means that the two histograms are drawn from the same distribution, in agreement with \citet{nava10}. We note that, \citet{bissaldi11} showed that the \ep distribution of some GBM-GRBs extends to higher energies compared to \batse \citep{kaneko06}. However, this is of no surprise as \citet{amati02} demonstrated that bursts with higher fluence, i.e. higher \eiso, have, on average, higher \ep (see also Sect.\ref{subsec:amati}). Since \citet{bissaldi11} only use bright GRBs, it is to be expected that their \ep distribution is shifted to higher energies.
We conclude that GBM, although being sensitive up to 40~MeV, does not find a previously undiscovered population of high-\ep GRBs, consistent with \citet{harris98}. 

The KS test was then applied to both the \ep distributions between the whole sample of GBM bursts and the 30 GBM bursts with measured redshift (for GRB 090519A and GRB 080928 a PL model fits the data best) and also to the \ep distribution of \batse bursts and the \ep distribution of GBM bursts with measured redshift. In neither case the differences were statistically meaningful ($P=24$\% and $P=20$\%, respectively). Thus, all 3 histograms are very likely drawn from the same distribution. In conclusion, the sample of GBM GRBs with measured redshift presented here is representative for the whole population of GRBs which were ever observed by \batse and GBM.

However, it should be stressed that GBM cannot measure \ep values which are lower than a certain limiting threshold. It is well known that \ep values of GRBs can go as low as a few keV. \citet{pel08} for example find \ep values as low as $\sim 2$~keV in GRBs that were observed by the High Energy Transient Explorer 2 (\emph{HETE-2}, see e.g. \citealt{hete2} and references therein). These low energetic events have been classified as X-ray flashes (XRF) or X-ray rich bursts (XRB) \citep[see e.g.][]{heise01, sakamoto05}. However, it is very likely that XRFs and XRBs are nothing else than weak and long GRBs \citep[see][and references therein]{kippen04}.

\begin{figure}[htbp]
\begin{center}
\includegraphics[width=0.5\textwidth]{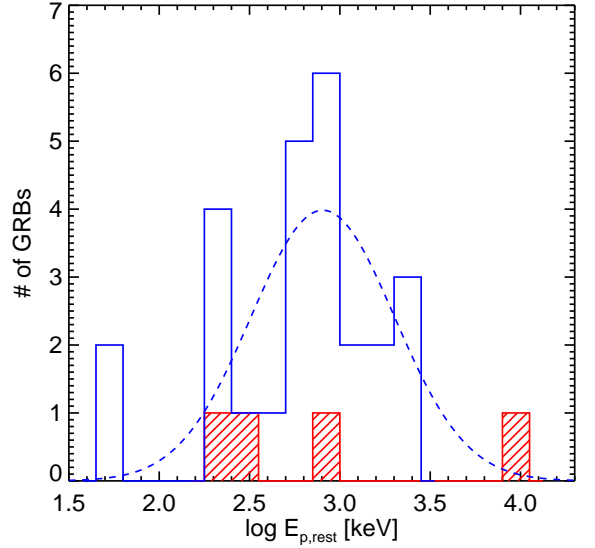}
\caption{Rest-frame distribution of \ep for 4 short (hatched histogram) and 26 long (empty histogram) GBM-GRBs. For two GRBs (GRB~090519A, GRB~080928) a simple PL model fits the data best, thus not providing an \ep value. The complete distribution has mean and median values of $1.1$~MeV and $750$~keV, respectively. A log-normal fit to the sample of long GRBs (dashed line) peaks at $\sim 800$~keV and has a $\rm{FWHM} = 0.93$ in log-space.}
\label{fig:histoeprest}
\end{center}
\end{figure}

The borderline \ep value is obviously located somewhere near the low-energy sensitivity of the NaIs, which has yet to be determined. Thus, in order to determine a potential bias in the \eprest distribution shown in Fig.~\ref{fig:histoeprest}, it is important to understand and quantify the limits of the GBM to measure \ep (be it either from the COMP model or Band function).

For this purpose, we created a set of simulated bursts with different initial spectral and temporal starting values. We input the source lifetime ($t_S$, 1~s, 5~s, 10~s, 100~s) and the photon flux ($f$) in the 10~keV to 1~MeV range (1, 3 and 10 ph cm$^{-2}$ s$^{-1}$). For the simulation the Band function was chosen as the photon model with varying \ep (15,17, 25, 50, 100 keV) but fixed $\alpha=-0.8$ and $\beta=-2.4$.  We simulate these bursts overlaid on real background data by using detector NaI~7 of GRB~090926A\footnote{The choice of NaI~7 and the choice for this specific GRB is completely arbitrary. We could have chosen any other detector that observed any other real GRB.}. This results in 60 different burst models. For each model, we created 1000 bursts to account for Poissonian noise. This results in 60000 spectra, each of which was then fitted with the Band function using the detector response matrix (DRM) of detector NaI~7 created for the location of GRB~090926A.

After the fitting procedure, we reject those bursts which have $\Delta  E_{\rm{p}} /{E_{\rm{p}}} \ge 0.3$ and $\sigma_\alpha \le 0.4$. These rejected spectra are then defined as unconstrained. We did not apply this criterion to the high-energy power law index $\beta$. A spectral fit that has a constrained \ep and $\alpha$ but an unconstrained $\beta$ is simply considered a COMP model.
In Table \ref{tab:sim} we report the mean and standard deviation of the output spectral parameters, \ep and $\alpha$, of the simulated bursts. 

The conclusions of this exercise are: 

\begin{enumerate}
\item GBM can recover \ep values only as low as $\approx15$~keV. The fact that the observed \ep distribution is indeed biased is in clear contradiction to e.g. \citet{brainerd00}. They argued that the observed \ep distribution (by \batse in this case) is actually the intrinsic one.

\item the input \ep parameter can be recovered from the simulated spectra within the 2~$\sigma$ errors for almost all simulated flux levels and source lifetimes.

\item GRBs with low fluxes, low \ep and short $t_s$ are more likely to be rejected than GRBs with higher fluxes, higher \ep or longer $t_s$

\item the low-energy power law index $\alpha$ tends to get softer, i.e. to have lower values, for low \ep and low fluxes. This last point is particularly noteworthy. \citet{crider97}, \citet{lloyd02} and later \citet{kaneko06} found a significant correlation between \ep and $\alpha$ in the time-resolved spectra of several GRBs. Supported by our simulations, we point out the possibility that a parameter reconstruction effect is at work in addition to any intrinsic correlation between \ep and $\alpha$. This effect has already been brought forward by \citet{preece98, lloyd00} and \citet{lloyd02}. In short, it depends on how quickly $\alpha$ can reach its asymptotic value and how close \ep is located to the low-energy limit of the instrument's energy bandpass. The closer \ep is situated to the detector's sensitivity limit, the fewer is the number of photons in the low-energy portion of the spectrum. This makes it increasingly difficult to determine the asymptotic value of $\alpha$. Instead, a more negative value of the low-energy power law index will be measured which is what is observed here.
\end{enumerate}

\renewcommand\arraystretch{1.15}
\begin{table}[ht!]
\caption{Mean and standard deviation of the output spectral parameters of the simulated bursts.}
\begin{center}
\begin{tabular}{lclccr}
\hline
 $t_S$ & $E_{\rm{p,in}}$ & $f$                                       & $E_{\rm{p,out}}$ & $\alpha_{out}$ & rej\\

 \hline
 \hline
 
 10 & 15 &10 &$ 15.0_{ -1.4}^{+  1.5}$ &$-1.53\pm0.36$ &$ 98$\\
100 & 15 & 1 &$ 21.7_{ -3.4}^{+  4.0}$ &$-1.55\pm0.16$ &$ 99$\\
100 & 15 & 3 &$ 15.4_{ -1.7}^{+  1.9}$ &$-1.39\pm0.30$ &$ 92$\\
100 & 15 &10 &$ 14.1_{ -0.5}^{+  0.5}$ &$-0.79\pm0.16$ &$ 46$\\
\hline
  1 & 17 & 3 &$ 22.6_{ -2.8}^{+  3.2}$ &$-1.48\pm0.25$ &$ 96$\\
  1 & 17 &10 &$ 23.5_{ -4.0}^{+  4.8}$ &$-1.48\pm0.20$ &$ 97$\\
  5 & 17 & 3 &$ 25.8_{ -4.7}^{+  5.8}$ &$-1.49\pm0.18$ &$ 99$\\
  5 & 17 &10 &$ 18.2_{ -2.1}^{+  2.4}$ &$-1.44\pm0.23$ &$ 88$\\
 10 & 17 & 3 &$ 21.2_{ -5.0}^{+  6.5}$ &$-1.55\pm0.17$ &$ 93$\\
 10 & 17 &10 &$ 17.6_{ -1.6}^{+  1.7}$ &$-1.26\pm0.27$ &$ 79$\\
100 & 17 & 1 &$ 22.5_{ -1.9}^{+  2.1}$ &$-1.33\pm0.23$ &$ 96$\\
100 & 17 & 3 &$ 16.9_{ -1.6}^{+  1.7}$ &$-1.28\pm0.25$ &$ 39$\\
100 & 17 &10 &$ 17.0_{ -0.4}^{+  0.4}$ &$-0.80\pm0.12$ &$  0$\\
\hline
  1 & 25 &10 &$ 30.6_{ -4.4}^{+  5.1}$ &$-1.30\pm0.28$ &$ 86$\\
  5 & 25 & 3 &$ 33.9_{ -5.6}^{+  6.8}$ &$-1.44\pm0.24$ &$ 96$\\
  5 & 25 &10 &$ 26.2_{ -1.9}^{+  2.0}$ &$-0.92\pm0.29$ &$ 39$\\
 10 & 25 & 3 &$ 30.3_{ -3.8}^{+  4.4}$ &$-1.29\pm0.27$ &$ 80$\\
 10 & 25 &10 &$ 25.4_{ -1.6}^{+  1.7}$ &$-0.89\pm0.25$ &$  4$\\
100 & 25 & 1 &$ 29.4_{ -3.3}^{+  3.7}$ &$-1.24\pm0.27$ &$ 73$\\
100 & 25 & 3 &$ 25.3_{ -1.5}^{+  1.6}$ &$-0.85\pm0.24$ &$  1$\\
100 & 25 &10 &$ 25.1_{ -0.6}^{+  0.6}$ &$-0.82\pm0.11$ &$  0$\\
\hline
  1 & 50 &10 &$ 53.3_{ -7.2}^{+  8.4}$ &$-0.88\pm0.28$ &$ 52$\\
  5 & 50 & 3 &$ 57.1_{ -7.8}^{+  9.0}$ &$-0.96\pm0.27$ &$ 62$\\
  5 & 50 &10 &$ 49.9_{ -5.2}^{+  5.8}$ &$-0.78\pm0.22$ &$  0$\\
 10 & 50 & 3 &$ 51.9_{ -6.4}^{+  7.3}$ &$-0.82\pm0.29$ &$ 37$\\
 10 & 50 &10 &$ 50.0_{ -3.8}^{+  4.1}$ &$-0.78\pm0.16$ &$  0$\\
100 & 50 & 1 &$ 50.3_{ -6.3}^{+  7.2}$ &$-0.80\pm0.28$ &$ 21$\\
100 & 50 & 3 &$ 49.9_{ -3.3}^{+  3.5}$ &$-0.78\pm0.14$ &$  0$\\
100 & 50 &10 &$ 50.0_{ -1.3}^{+  1.3}$ &$-0.80\pm0.05$ &$  0$\\
\hline
  1 &100 & 3 &$121.0_{ -5.2}^{+  5.4}$ &$-0.82\pm0.13$ &$ 99$\\
  1 &100 &10 &$ 92.0_{-15.7}^{+ 18.9}$ &$-0.71\pm0.25$ &$ 16$\\
  5 &100 & 3 &$ 98.7_{-17.7}^{+ 21.5}$ &$-0.74\pm0.26$ &$ 21$\\
  5 &100 &10 &$ 98.6_{-10.1}^{+ 11.2}$ &$-0.78\pm0.13$ &$  0$\\
  10 &100 & 3 &$ 95.0_{-16.0}^{+ 19.3}$ &$-0.73\pm0.23$ &$  4$\\
10 &100 &10 &$ 99.0_{ -7.7}^{+  8.4}$ &$-0.79\pm0.09$ &$  0$\\
100 &100 & 1 &$ 92.4_{-14.6}^{+ 17.3}$ &$-0.71\pm0.23$ &$  8$\\
100 &100 & 3 &$ 99.9_{ -6.6}^{+  7.1}$ &$-0.80\pm0.08$ &$  0$\\
100 &100 &10 &$ 99.9_{ -2.4}^{+  2.4}$ &$-0.80\pm0.03$ &$  0$\\

 \hline
\end{tabular}
\tablefoot{
This Table shows the input source lifetime, $t_S$ (in $\rm{s}$), $E_{\rm{p,in}}$ (in keV), photon flux, $f$ (in ph cm$^{-2}$~s$^{-1}$), the recovered \ep, $E_{\rm{p,out}}$ (in keV), the recovered low-energy PL index, $\alpha$, and the percentage of bursts which were rejected due to the selection cut, rej (in \%). The input low-energy power law index was $-0.8$. Cases for which all simulated bursts were rejected are not reported in this Table (for details see text). \\
 }
\end{center}
\label{tab:sim}

\end{table}%

\subsubsection{The intrinsic \ep distribution}

In Fig.~\ref{fig:histoeprest} we present a histogram of \eprest of our sample of GBM GRBs where a \ep measurement was possible. The distribution of all bursts has mean and median values of $1.1$~MeV and $750$~keV, respectively. A log-normal fit to the sample of long bursts peaks at $\sim 800$~keV. 
In the canonical scenario of GRB jets, turbulent magnetic fields build
up behind the internal shock, and electrons produce a synchrotron power law
spectrum. The typical rest-frame frequency $\nu_{\rm rest}$ in the internal shock dissipation is 
$\nu_{\rm rest}\propto 0.2\: L_{52}^{1/2}\: r_{13}^{-1} {\rm MeV}$ \citep{zhang02}, 
where $L_{52}$ is the luminosity in 10$^{52}$~erg/s and
$r_{13}$ the dissipation radius in units of 10$^{13}$~cm. Our measurement of
the rest-frame peak energy therefore leads to a constraint of 
$L_{52}^{1/2}\:r_{13}^{-1} \approx 0.8$ (in the above units).

Recently, \citet{collazzi11} reported that the width of the \eprest distribution must be close to zero with the peak value located close to the rest-mass energy of electrons at 511~keV. This effectively implies that all GRBs must be thermostated by some unknown physical mechanism. We tested this claim by determining the width of the \eprest distribution ($\sigma_{E_p (1+z)}=0.48$) together with all the individual errors that add to the uncertainty of the \ep measurement in log-space (for details refer to eq.~1 in \citealt{collazzi11}):

\begin{equation}
\sigma_{\rm{Total}}^2 = \sigma_{\rm{Poisson}}^2 + \sigma_{\rm{Det}}^2 + \sigma_{\rm{Choice}}^2 + \sigma_{\rm{Def}}^2.
\end{equation}

$\sigma_{\rm{Choice}}$ describes the different choices made by different analysts. Since all the bursts in this paper were analyzed consistently (selection of the time interval, energy range, etc.), we can set $\sigma_{\rm{Choice}} =0$. $\sigma_{\rm{Det}}$ is the error which results in not knowing the detector response perfectly. However, according to \citeauthor{collazzi11}, $\sigma_{\rm{Det}}$ is negligibly small, meaning that the calibrations of the detectors are usually well understood. $\sigma_{\rm{Def}}$ describes the differences that are obtained in \ep when using different photon models. As it was shown already by e.g. \citet{band93a} and \citet{kaneko06} the COMP model results in higher \ep values than the Band function. \citeauthor{collazzi11} use $\sigma_{\rm{Model}}=0.12$ to account for this difference. Here, this value is adopted as $\sigma_{\rm{Def}}$. The Poissonian errors are $\sigma_{\rm{Poisson}} \approx 0.10$ in log-space.
Thus $\sigma_{\rm{Total}}=\sqrt{0.12^2+0.1^2}=0.16$.

Inserting the just found values in eq.~8 in \citeauthor{collazzi11} gives

\begin{equation}
\sigma_{E_p\rm{int}}^2 = \sigma_{E_p (1+z)}^2-\sigma_{\rm{Total}}^2 = 0.45.
\end{equation}

However, a zero width in the \ep distribution is synonymous with $\sigma_{E_p\rm{int}}=0$. The conclusion is that the \eprest distribution does not have a zero width. This finding is in conflict with the implications and conclusions discussed in \citeauthor{collazzi11}

\subsection{GRB duration (\tninety)}
For determining the duration of a GRB, we applied the method first introduced by \citet{kouv93}. 
They defined the burst duration as the time in which 90\% of the burst counts is collected (\tninety). Here, we adopted the same definition; however, the burst's fluence was used instead of the counts.
The \tninety value depends highly on the detector and on the energy interval in which it is determined \citep[see e.g.][]{bissaldi11}. 
Additionally, since we are interested in durations in the rest-frame of the GRB, it is not sufficient to simply account for the time dilation due to cosmic expansion by dividing the measured durations
by $(1+z)$. The energy band in which the \tninety is determined needs to be redshift corrected as well.
We determine the burst duration in fluence space in the rest-frame energy interval from  $50$~keV to $300$~keV, i.e. in the observer frame energy interval from $50/(1+z)$~keV to $300/(1+z)$~keV.

\subsubsection{The intrinsic \tninety distribution }
\begin{figure}[htbp]
\begin{center}
\includegraphics[width=0.5\textwidth]{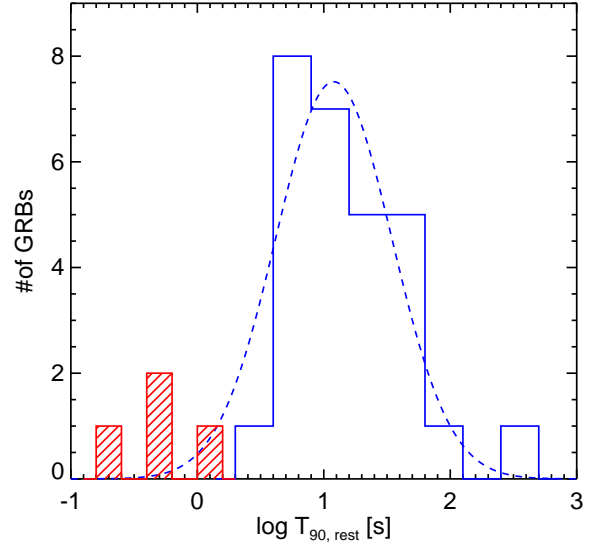}
\caption{\tninetyrest distribution of 4 short (red hatched histogram) and 28 long (blue histogram) GBM GRBs in the redshift corrected energy interval from 50~keV to 300~keV. The log-normal fit to the distribution (dashed line) of the long bursts peaks at $\sim 12$~s and has a $\rm{FWHM}=1.1$ in log-space.}
\label{fig:histot90}
\end{center}
\end{figure}

In Fig.\ref{fig:histot90} we present the rest-frame distribution of \tninety of our sample of 32 GBM GRBs. The number of short bursts is still too small to unambiguously recover a bimodal distinction of short and long events in the rest-frame. Henceforth, a distinction is made between the short and long class of GRBs using an ad-hoc definition of $T_{90,\rm{rest}} \le 2$~s for short bursts and $T_{90,\rm{rest}} > 2$~s for long bursts, although we do have neither observational nor physical evidence to support such a distinction in the GRB sample presented here. All GRBs in our sample that are defined as short in the observer-frame, also remain short with the here adopted definition. We note that GRB~100816A is peculiar in that it is classified as a short burst by GBM data, but it has a $T_{90}=2.9 \pm 0.6$ in the $15$~keV - $350$~keV energy range in \emph{Swift} data with a low-level emission out to about T0+100 \citep{mark10}.

The mean value of the whole \tninetyrest distribution is $32.4$~s  and the median value is $10.8$~s.

\subsection{Isotropic energy (\eiso)}
\begin{figure}[htbp]
\begin{center}
\includegraphics[width=0.5\textwidth]{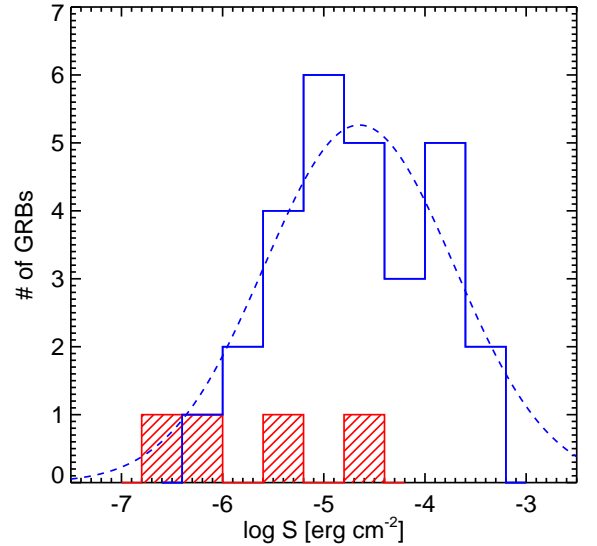}
\caption{Energy fluence distribution of 32 GBM GRBs determined in the energy range between $1/(1+z)$~keV to $10/(1+z)$~MeV range. The complete distribution has mean and median values of $1.6\times10^{-5} \rm{erg}\, \rm{cm}^{-2}$ and $5.9\times 10^{-5} \rm{erg}\, \rm{cm}^{-2}$. A log-normal fit (dashed line) to the long GRBs peaks at $2.2\times10^{-5} \rm{erg}\, \rm{cm}^{-2}$ and has a $\rm{FWHM}=2.2$ in log-space.}
\label{fig:flugbm}
\end{center}
\end{figure}

\eiso is calculated by 

\begin{equation}
E_{\rm{iso}} = \frac{4 \pi d_L^2 }{1+z} \, S_{\gamma} , 
\end{equation}

where $d_L$ is the luminosity distance and $S_\gamma$ the fluence in the $1/(1+z)$~keV to $10/(1+z)$~MeV frame. We determine $S_\gamma$ using the energy flux provided by the best-fit spectral parameters and multiplying it with the total time interval over which the fit was performed. Since we performed the fit for time intervals where the count rate exceeded a S/N ratio of 3.5, it happened that some time intervals of some bursts were not included in the fit (e.g. phases of quiescence where the count rate dropped back to the background level). These time intervals were not used to calculate the fluence. The $S_\gamma$ distribution is shown in Fig.\ref{fig:flugbm}. The median value of the fluence distribution is $1.6\times10^{-5}$~erg~cm$^{-2}$ and the mean value is $5.9\times10^{-5}$~erg~cm$^{-2}$. A log-normal fit to the data peaks at $2.2\times10^{-5}$~erg~cm$^{-2}$.

\begin{figure}[htbp]
\begin{center}
\includegraphics[width=0.5\textwidth]{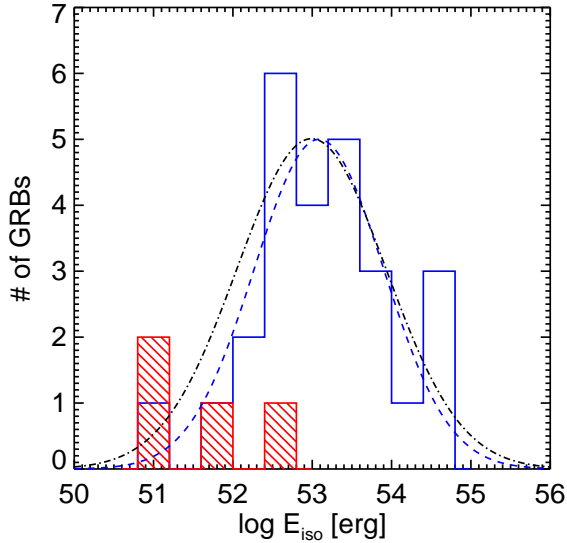}
\caption{\eiso distribution for long (blue hatched histogram) and short (red hatched histogram). For long bursts the distribution has mean and median values of $1.4 \times 10^{53}$~erg and $1.2 \times10^{53}$~erg, respectively. A log-normal fit to the long bursts (dashed line) peaks at $1.2\times 10^{53}$~erg with a $\rm{FWHM}=1.9$ in log-space. Short bursts have a mean and median value of $4 \times 10^{51}$~erg and $2.9 \times 10^{51}$~erg, respectively. A log-normal fit to the combined sample (dash-dotted line) peaks at $9.8 \times 10^{52}$~erg with a $\rm{FWHM}=2.2$ in log-space. }
\label{fig:histeiso}
\end{center}
\end{figure}

The \eiso distribution is shown in Fig.\ref{fig:histeiso}. The distribution for the long bursts has a median and mean value of $1.2\times10^{53}$~erg and $1.4\times10^{53}$~erg, respectively. Short bursts, on the other hand, have significantly lower values of $2.9\times10^{51}$~erg and $4.0\times10^{51}$~erg, respectively. A log-normal fit to the long bursts reveals a central value of $10^{53.1}$~erg. Because our sample is dominated by long GRBs a log-normal fit to the whole distribution results in an essentially unchanged peak value ($10^{53}$~erg).

\section{Correlations}
\subsection{Amati relation}
\label{subsec:amati}
\begin{figure}[htbp]
\begin{center}
\includegraphics[width=0.5\textwidth]{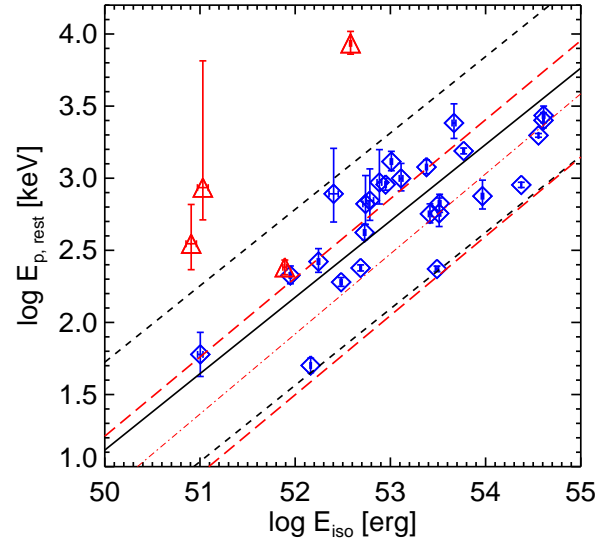}
\caption{Amati relation for 4 short (red open triangles) and 26 long (blue open diamonds) GBM GRBs. Also shown is the best power-law fit to the data (black solid line) and the extrinsic $2\sigma$ scatter (black dashed lines) with the best power-law fit published by \citet{amati10} (red dash-dotted line) with the $2\sigma$ scatter (red long dashed line).}
\label{fig:amati}
\end{center}
\end{figure}

\citet{amati02} first showed that there is a tight correlation between \eprest and \eiso (the isotropic equivalent bolometric energy determined in the energy range between 1~keV to 10~MeV). This relation is now known as the ``Amati relation''. In Fig.\ref{fig:amati} we show the Amati relation for the 30 GBM GRBs with measured \eprest and \eiso. While there is an evident correlation between these two quantities (Spearman's rank correlation of $\rho= 0.74$ with a chance probability of $1.7\times10^{-5}$) the extrinsic scatter of the long GRBs is larger by a factor of $\sim 2$ in $\log$-space compared to \citet{amati10}. Also, the best fit to our data is shifted to slightly larger \eprest values. 
The best fit power-law index to the long GRBs of our sample is $0.52\pm0.06$ which is in agreement with the indices obtained by e.g. \citet{amati10, ghina09, ghirlanda10}. As has been shown by other authors in the past \citep[see e.g.][]{amati08, ghina09, amati10} short bursts do not follow the relation, being situated well outside the 2~$\sigma$ scatter around the best-fit. This is true also for the power-law fit derived here (see Fig.\ref{fig:amati}) except for GRB~100816A. However, as already stated above this burst may actually fall in an intermediate or hybrid class of short GRBs with extended emission \citep[see e.g.][]{norris06, zhang09}. 

\subsection{Yonetoku relation}
\citet{yonetoku04} found a tight correlation between the rest frame peak energy in the $\nu \rm{F}_{\nu}$ spectrum \eprest  and the 1-s peak luminosity $(L_p)$ in GRBs (so called Yonetoku relation). The peak luminosity is calculated with 

\begin{equation}
L_p=4\pi d_l^2 F_p,
\end{equation}

where $d_l$ is the luminosity distance and $F_p$ the 1~s peak energy flux. We determine $F_p$ in the energy range between 30 keV and 10 MeV in the rest frame of the GRB \citep{yonetoku04}. We use the GBM peak spectral catalogue (Goldstein et al. 2011) to determine the best fit spectral parameters of the brightest 1~s bin. We then determined the energy flux in the rest-frame of the GRB by integrating 

\begin{equation}
F = \int_{30/(1+z)}^{10000/(1+z)} N_0 \; E \; \Phi(E) \; dE,
\end{equation}

where $N_0$ is the normalization of the spectrum (in photons cm$^2$ s$^{-1}$), $E$ the energy and $\Phi(E)$ the form of the spectrum (either Band or COMP). In order to determine the error on the energy flux, we calculate 1000 flux values for each GRB by varying the input parameters, $N_0$, \ep, $\alpha$ and $\beta$ according to the $1 \sigma$ error on each parameter. We then use the median and the mean absolute deviation (MAD) around the median of the 1000 simulated bursts to determine the peak energy flux and the error on the latter, respectively.

\begin{figure}[htbp]
\begin{center}
\includegraphics[width=0.5\textwidth]{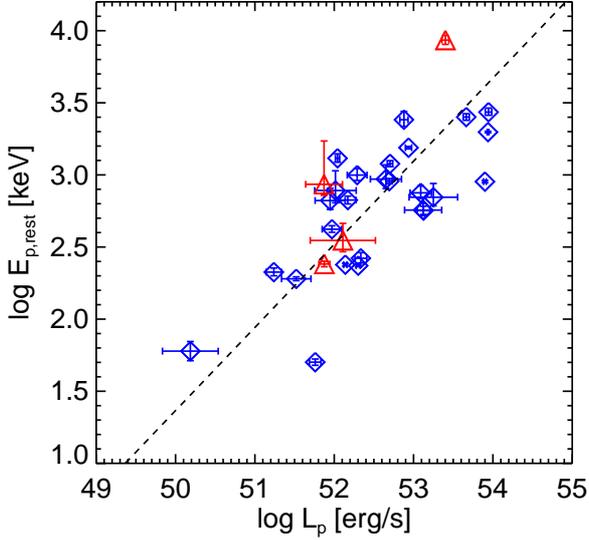}
\caption{Yonetoku relation for 4 short (red open triangles) and 26 long (blue open diamonds) GBM GRBs. Also shown is the best power-law fit to the data (black dashed line).}
\label{fig:yonetoku}
\end{center}
\end{figure}

We present this relation for 30 GBM GRBs in Fig.~\ref{fig:yonetoku} and find a best-fit power law index of $0.58 \pm 0.08$. The Spearman's rank correlation gives $\rho=0.7$ with a chance probability of $2.3\times10^{-5}$. Our findings are in good agreement with \citet{yonetoku04, ghina09} and \citet{ghirlanda10}.

\subsection{\tninetyrest vs redshift}
\begin{figure}[htbp]
\begin{center}
\includegraphics[width=0.5\textwidth]{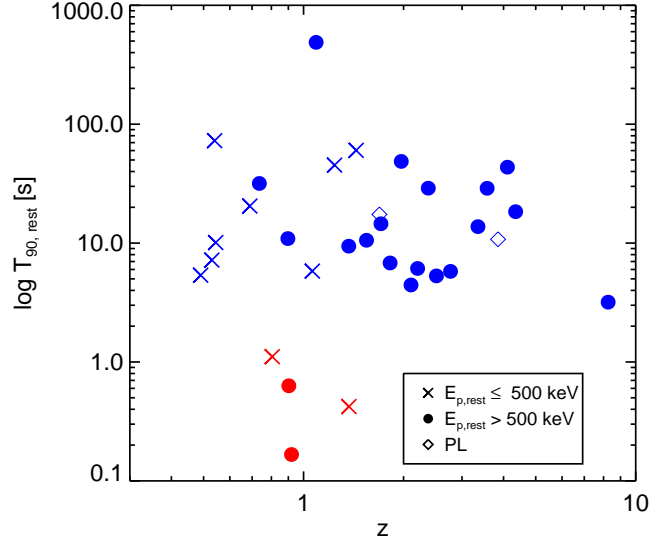}
\caption{Testing the cosmic evolution of \tninetyrest. No correlation is evident. Bursts with $E_{p,\rm{rest}} \le 500$~keV are shown as crosses, bursts with  $E_{p,\rm{rest}} > 500$~keV shown as filled circles. The two bursts for which are best fit by a simple power-law are shown as open diamonds.}
\label{fig:t90vsz}
\end{center}
\end{figure}

The rest-frame \tninety is plotted as a function of redshift in Fig.~\ref{fig:t90vsz}.
Contrary to \citet{pel08}, who find a negative correlation between \tninetyrest and $z$, we do not find any evidence in the GBM data for such a correlation. Also, there is no correlation between \tninetyrest and $z$ when accounting for \eprest. A Spearman rank test for all bursts gives a correlation coefficient of $\rho=0.04$ with a chance probability of $P=0.81$. However, the p-value in \citeauthor{pel08} is $\approx 10^{-3}$ which makes it a weak case for such a correlation in the first place. Our results confirm the analyses with \emph{Swift} detected GRBs \citep{greiner11}. We do note a lack of short GRBs for $z \ge 2$. This, however, is very likely a selection bias, because firstly, short GRBs at such high redshifts must be very luminous to be observed by GBM and secondly, short GRBs are subluminous in the optical band \citep{kann08} and therefore it is difficult to obtain a redshift measurement.

\subsection{\eprest vs redshift }
In order to explain the detection rate of GRBs at high-$z$, \citet{salva07} conclude that high-$z$ GRBs must be more common \citep[e.g.][]{daigne06,wang11} and/or intrinsically more luminous \citep{salva09} than bursts at low-$z$ \citep[but see][]{but10}. As already mentioned above, \citet{yonetoku04} found a tight correlation between the 1-s peak-luminosity ($L_p$) and \eprest in GRBs. Assuming that the luminosity function of GRBs indeed evolves with redshift and that the Yonetoku relation is valid, we would also expect a positive correlation of \eprest with $z$.

\begin{figure}[htbp]
\begin{center}
\includegraphics[width=0.5\textwidth]{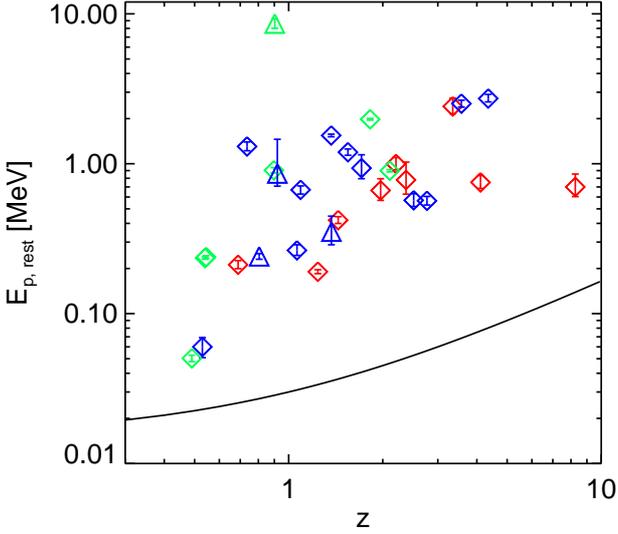}
\caption{Testing the cosmic evolution of \eprest for long (diamonds) and short (triangles) GRBs. The solid line indicates the redshift corrected lower limit for GBM to measure \ep which is currently estimated to be $\sim 15$~keV in the observer frame (see Table~\ref{tab:sim}). Bursts with high (green), intermediate (blue) and low (red) peak photon flux (in ph~cm$^{-2}$~s$^{-1}$) are labeled.}
\label{fig:epvsz}
\end{center}
\end{figure}

In Fig.\ref{fig:epvsz} we present \eprest vs $z$. As was shown in Sect.\ref{subsubsec:instbias} and in Table~\ref{tab:sim}, GBM can reliably measure \ep down to $\sim 15$~keV. The solid line indicates this redshift-corrected lower limit. The Spearman's rank correlation, using only the long GRBs, is $\rho=0.58$ with a chance probability of $P=2\times10^{-3}$. When including the short GRBs, the correlation coefficient effectively remains unchanged, whereas the chance probability increases to $P=4\times10^{-3}$, making a correlation slightly less likely.

However, this correlation can be explained entirely by selection effects: GRBs do not populate the empty area in Fig.~\ref{fig:epvsz} (low \eprest and $z>1$) because they simply can not be detected by GBM. Even though GBM \textit{could} recover a low \ep value of such GRBs, as was shown above, the detection of such events is very challenging because of the low photon fluxes of these events. As one can see in Fig.\ref{fig:epvsz} the lower boundary of the apparent correlation is composed of the bursts that have relatively low peak photon fluxes. This is already an indication that these events reside at the lower fluence limit for GBM to both trigger on these events.

Since we actually know the intrinsic parameters of the 32 GRBs, one can test up to which maximum redshift, $z_{\rm{max}}$ these bursts could have been detected, i.e. for which GBM would have triggered. GBM has many trigger algorithms (various trigger time scales for various energy ranges). For the purpose of this test, we focus on the $50$~keV to $300$~range which is the classical trigger energy range for a GRB and a timescale of a maximum of $4.096$~s for long GRBs and $1.024$~s for short GRBs. In order to shift a GRB to a higher redshift, three observables change: 

\begin{enumerate}
\item The duration $T_{90} (z_{\rm{max}}) = T_{90} (z_0) \frac{1+z_{\rm{max}}}{1+z_0}$
\item The peak energy $E_{\rm{p}} (z_{\rm{max}}) = E_{\rm{p}} (z_0) \frac{1+z_0}{1+z_{\rm{max}}}$
\item The flux of the GRB.
\end{enumerate}

While it is straightforward to account for the changes of \ep and \tninety, the proper treatment of the flux is more complex. RMFIT outputs the spectral parameters, including the normalization ($N_0$ in ph~cm$^{-2}$~s$^{-1}$~keV$^{-1}$) of the spectrum, which can be recognized as a proxy for the flux of a GRB. Therefore, in order to decrease the flux when shifting the GRB to ever higher redshifts, the normalization has to be decreased accordingly.
This was done as follows:

The photon luminosity of a GRB, in the energy band from $E_1$ to $E_2$ is defined as follows

\begin{equation}
L = 4 \pi d_{L,z_0}^2 \int_{E_1/(1+z_0)}^{E_2/(1+z_0)} N_0 \Phi_1(E)\,dE, 
\end{equation}

where $d_L$ is the luminosity distance and $z_0$ the redshift of the burst. The integral describes the photon flux of the burst in the $E_1$ to $E_2$ energy range at the rest-frame of the burst. $N_0$ is the normalization of the spectrum and $\Phi_1(E)$ is the shape of the spectrum (Band or COMP) with $E_p = E_p(z_0)$, i.e. $\Phi_2=\Phi_2(E,E_p(z_0))$.
The luminosity of a burst is independent of redshift. This, in turn, means that the above equation is valid also when exchanging $z_0$ with $z_{\rm{max}}$. One gets 

\begin{equation}
L = 4 \pi d_{L,z_{\rm{max}}}^2  \int_{E_1/(1+z_{\rm{max}})}^{E_2/(1+z_{\rm{max}})} N_{z_{\rm{max}}} \Phi_2(E)\,dE,
\end{equation}

where $\Phi_2=\Phi_2(E,E_p(z_{\rm max}))$.

Setting the two equations equal and solving for $N_{z_{\rm{max}}}$ one is left with

\begin{equation}
N_{z_{\rm{max}}} = N_0 \times \frac{d_{L,z_0}^2}{d_{L,z_{\rm{max}}}^2} \times \frac{  \int_{E_1/(1+z_0)}^{E_2/(1+z_0)} \Phi_1(E,E_p(z_0))\,dE}{\int_{E_1/(1+z_{\rm{max}})}^{E_2/(1+{z_{\rm{max}}})} \Phi_2(E,E_p(z_{\rm max}))\,dE}.
\end{equation}

It can be shown, for both the COMP model and the Band function, that the fraction of the two integrals is nothing else than 

\begin{equation}
\frac{  \int_{E_1/(1+z_0)}^{E_2/(1+z_0)} \Phi_1(E,E_p(z_0))\,dE}{\int_{E_1/(1+z_{\rm{max}})}^{E_2/(1+{z_{\rm{max}}})} \Phi_2(E,E_p(z_{\rm max}))\,dE} = \frac{(1+z_{\rm max})^{\alpha+1}}{(1+z)^{\alpha +1}},
\end{equation}

 where $\alpha$ is the low-energy power law index.
 
 Thus,
\begin{equation}
N_{z_{\rm{max}}} = N_0 \times \frac{d_{L,z_0}^2}{d_{L,z_{\rm{max}}}^2} \times \frac{(1+z_{\rm max})^{\alpha+1}}{(1+z)^{\alpha +1}}.
\end{equation}

\begin{figure}[htbp]
\begin{center}
\includegraphics[width=0.5\textwidth]{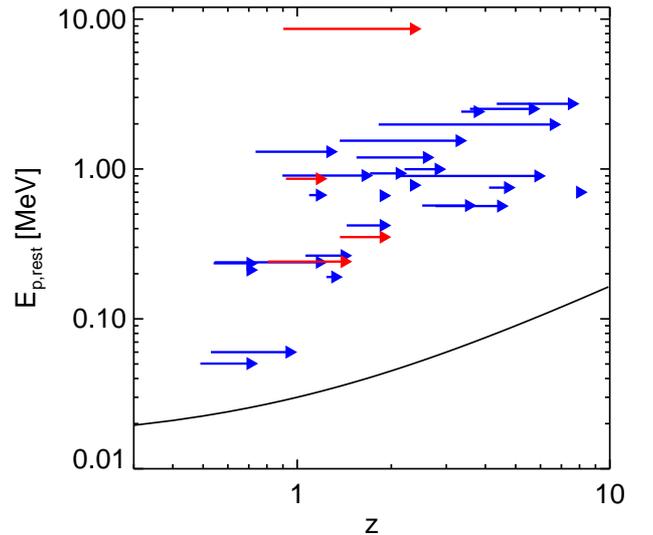}
\caption{Same as Fig.~\ref{fig:epvsz} but with a line indicating the maximum possible redshift, $z_{\rm{max}}$ that these 30 GRBs can have to still be detectable by GBM.}
\label{fig:epvszmax}
\end{center}
\end{figure}

For GBM to trigger, two detectors need to be above the trigger threshold. Therefore, real background information of the second brightest detector of every burst was used. With this background data, we shift each bursts in steps of $\Delta z = 0.25$ to higher redshifts and simulate 1000 bursts for each redshift step  (changing the source lifetime, input \ep and normalization as described above, additionally adding Poissonian noise to the best-fit parameters) by forward folding the photon model through the detector response matrix, created at the time and location of the real GRBs. To determine if GBM would have triggered, we determine the signal-to-noise ratio (SNR) with

\begin{equation}
\rm{SNR}=\frac{\Delta t \cdot c_s- \Delta t\cdot c_b}{\sqrt{\Delta t \cdot c_b}},
\end{equation}

where, $\Delta t$ is the trigger time scale ($\Delta t=4.096$~s for long GRBs and $\Delta t=1.024$~s for short GRBs), $c_s$ and $c_b$ are the counts/s of the source and background, respectively. If more than $90\%$ of the simulated bursts have $\rm{SNR} \ge 4.5$ it can safely be assumed that the burst would have been detected at this redshift. We plot \eprest, which obviously remains constant at all redshifts, vs the range of the actually measured to maximum redshift in Fig.\ref{fig:epvszmax}.  
We already know that the here presented sample of GRBs is representative of all bursts detected by GBM (see Sect.~\ref{subsubsec:instbias}). Therefore, we note that the determination of the lowest measurable \ep value is not as crucial as the determination of the detector sensitivity. None of the bursts in our sample populates the empty region at $2 \le z \le 8$ and $50 \le E_{p,\rm{rest}} \;\rm{[keV]}\le 500$ even when shifting them at their maximum detectable redshift. We conclude that the correlation between \eprest and $z$ can be explained entirely by the sensitivity limitations of the instrument. 

\subsubsection{Low number statistics?}
Another way to test if the $E_p - z$ correlation is genuine is by bearing in mind that it arises simply due to the lack of GRBs with measured redshift and $E_p < 100$~keV. In the present redshift sample there are only 4 GRBs with such a low \ep value. Considering the \ep distribution for 488 GBM bursts of the first two years of operations (Fig.~\ref{fig:epbatse}) it is possible to estimate as to how many bursts with $E_p < 100$~keV are expected if 32 GRBs are drawn randomly from the full GBM sample. After a run of $10^4$ drawings, a distribution is obtained which peaks at $\sim 7$. This means that, on average, 7 out of 32 GRBs are expected to have $E_p<100$~keV. According to our simulation the probability of observing only 4 such GRBs, as we do in our actual sample, is  $P=8\%$, which is within the $2\sigma$ limit and thus not significant. Therefore, more GRBs with redshift measurement are required to understand if the ``blank area'' at low \ep values between $2 < z < 5$ in Fig.~\ref{fig:epvsz} is simply underrepresented or not populated at all.

\section{Summary \& Conclusions}

The \emph{Fermi}/GBM is a key instrument to study the temporal and spectral properties of GRBs. For bursts which have redshift measurements it becomes even more important since it allows GRBs to be studied in their rest-frame. Here, we presented such a study for 32 GRBs observed by GBM, focusing on both the temporal and spectral properties, as well as on intra-parameter relations within these quantities.

The \eprest distribution of the GBM GRB sample covers an energy range from tens of keV up to several MeV, peaking at $\approx 800$~keV. Despite the broader energy coverage of GBM compared to \batse, no high-\ep population is found. However, the GBM \ep distribution is strongly biased against XRFs or XRBs which have \ep values that fall below 15~keV.  Additionally, we confirm a previously reported parameter reconstruction effect, namely that the low-energy power law index $\alpha$ tends to get softer when \ep is close to the lower end of the detector energy range.

Using the canonical internal shock model for GRBs, the mean \eprest ($\approx 800$~keV) implies a dissipation radius of $r_{13}\approx 1.3\:  L_{52}^{1/2}$~cm.

Another finding of this work is that the width of the \ep distribution is, in fact, not close to zero. Such a claim has recently been brought forward by \citet{collazzi11} who argue that all GRBs are thermostated and thus, have the same \eprest. If true, this would require an unknown physical mechanism that holds GRBs at a constant \eprest value. However, it could be shown here that such a mechanism is not required because our sample has an \eprest distribution that ranges over several decades in energy.

We find that the \tninetyrest stretches from tenths of a second to several hundreds of seconds with a median value of $\approx 10$~s.

The \eiso distribution ranges from $10^{51}$~erg to $10^{55}$~erg, peaking at $\approx 10^{53}$~erg.
We confirm the \eprest - \eiso correlation and find a power law index of $0.5$, consistent with the values reported in the literature but with a significantly larger scatter around the best-fit. We also confirm a strong correlation between the 1~s peak luminosity of the burst and its \eprest with a best-fit power law index of $0.57$.
 
We looked for additional correlations between the parameters. We did not find any evidence for a cosmic evolution of \tninetyrest. Although a correlation between \eprest and $z$ is not entirely unexpected from theoretical considerations and looks intriguing on a \ep - $z$ plot, we conclude that the apparent relationship is simply arising due to the detector sensitivity.

\acknowledgements{We thank Jonathan Granot for useful discussions. AJvdH was supported by NASA grant NNH07ZDA001-GLAST. SF acknowledges the support of the Irish Research Council for Science, Engineering and Technology, cofunded by Marie Curie Actions under FP7.The GBM project is supported by the German Bundesministerium f\"ur Wirtschaft und Technologie (BMWi) via the Deutsches Zentrum f\"ur Luft- und Raumfahrt (DLR) under the contract numbers 50 QV 0301 and 50 OG 0502.}

\bibliographystyle{aa}
\bibliography{references}
\clearpage
\renewcommand\arraystretch{1.3}
\begin{table*}[htbp]
\caption{Rest-frame parameters of 32 GBM GRBs.}             
\label{table:2}      
\centering          
\begin{tabular}{lllllll} 
\hline
GRB 		&Mission &$T_{90,\rm{rest}}$	 		&\eprest 		&Best Fitting 		&$z$ 		&Ref.  \\
    		&        &[s]             		& 	[keV] 	 	&     		Model							& 		&       \\
\hline
\hline
100816A 	&Swift   & $  1.11\pm 0.11$	& $  241_{  -11}^{   +11}$  	&BAND   	 	& 0.8049(1) & [1]    								\\
100814A 	&Swift   & $ 60.25\pm 0.90$	& $  420_{  -21}^{   +23}$  	&COMP  		 	& 1.44(1)   & [2]    							\\
100414A 	&Swift   & $  9.42\pm 0.21$	& $ 1544_{  -31}^{   +31}$	&COMP 				& 1.368(1)  & [3]      							\\
100117   	&Swift   & $  0.17\pm 0.07$	& $  861_{ -152}^{  +595}$	&COMP 				& 0.92(1)   & [4]    								\\
091208B 	&Swift   & $  5.82\pm 0.19$	& $  264_{  -20}^{   +24}$	&COMP  				& 1.063(1)  & [5]    							\\
 091127  	&Swift   & $  5.37\pm 0.20$	& $  50_{   -3}^{    +3}$  	&BAND    		& 0.490(1)  & [6]     						\\
091024\tablefootmark{ a} 		&Swift	 & $ \approx 487.80 $	& $  659 _{  -67}^{   +90}$	&COMP				& 1.092(1)  & [7]	 								\\
091020  	&Swift   &  $ 14.54\pm 1.18$	& $  935_{ -140}^{  +213}$	&BAND   			& 1.71(1)   & [8]     							\\
 091003A 	&Fermi 	 &   $ 10.91\pm 0.16$	&  $  904_{  -36}^{   +39}$	&COMP  			&0.8969(1)  & [9]     							\\
 090927  	&Swift   &  $  0.42\pm 0.04$	& $  351_{   63}^{   +96}$	&COMP    			&1.37(1)    & [10]    							\\
 090926B 	&Swift   &   $ 45.27\pm 2.05$	&  $  190_{   -6}^{   + 6}$	&COMP    		&1.24(1)    & [11]    							\\
 090926A 	&Fermi   &   $  4.44\pm 0.06$	&  $  898_{  -16}^{   +17}$	&BAND*   		&2.1062(1)  & [12]    							\\
 090902B 	&Fermi   &   $  6.80\pm 0.11$	&  $ 1980_{  -25}^{   +26}$	&COMP+PL 		&1.822(1)   & [13]    							\\
 090618  	&Swift   &  $ 72.66\pm 0.45$	& $  235_{   -4}^{    +5}$	&BAND*   			&0.54(1)    & [14] 	   							\\
 090519A 	&Swift   &   $ 10.76\pm 4.06$	&\nodata 			  	&PL      		&3.85(1)    & [15, 16]   				\\
 090516A 	&Swift   &   $ 43.49\pm 3.70$	&   $  751_{  -67}^{   +83}$	&COMP    	 	&4.109(1)   & [17, 18]    				\\
 090510  	&Swift   &  $  0.63\pm 0.11$	&  $ 8605_{ -592}^{  +660}$ 	&BAND+PL 	   	&0.903(1)   & [19]     							\\
 090424  	&Swift   &  $ 10.10\pm 0.39$	&  $  239_{   -5}^{    +5}$	&BAND    		&0.544(1)   & [20, 21]     				\\
 090423  	&Swift   &  $  3.19\pm 0.16$	&  $  700_{  -96}^{  +154}$ 	&COMP    	 	&8.26(8)    & [22]      	\\
 090328A 	&Fermi   &   $ 31.74\pm 0.58$	&   $ 1304_{  -83}^{   +93}$	&COMP    		&0.736(1)   & [23]   							\\
 090323  	&Fermi   &  $ 28.91\pm 0.13$	&  $ 2518_{ -131}^{  +135}$	&BAND    		&3.57(1)    & [24]   							\\
 090102  	&Swift   &  $ 10.56\pm 0.27$	&  $ 1194_{  -55}^{   +60}$	&COMP    		&1.547(1)   & [25]   							\\
 081222  	&Swift   &  $  5.78\pm 0.48$	&  $  566_{  -35}^{   +39}$	&BAND    		&2.77(1)    & [26]   							\\
 081121  	&Swift   &  $  5.30\pm 0.46$	&  $  570_{  -52}^{   +61}$ 	&BAND     	     	&2.512(1)   & [27]   							\\
 081008  	&Swift   &  $ 48.64\pm 6.80$	&  $  664_{  -94}^{  +130}$ 	&COMP    	    	&1.9685(1)  & [28]   							\\
 081007  	&Swift   &  $  7.19\pm 1.31$	&  $   60_{   -9}^{    +9}$ 	&COMP    	     	&0.5295(1)  & [29]   							\\
 080928		&Swift	 &  $ 17.42\pm 2.30$	& \nodata				&PL			&1.692(1)   & [30] 							\\
 080916C	&Fermi   &   $ 18.37\pm 2.13$	&  $ 2724_{ -138}^{  +177}$ 	&BAND     	    	&4.35(5)    & [31] 							\\
080916A 	&Swift   &   $ 20.49\pm 0.36$	&  $  212_{  -12}^{   +14}$ 	&COMP    	      	&0.689(1)   & [32]   							\\
 080905B 	&Swift   &   $ 28.96\pm 0.59$	&  $  779_{ -153}^{  +246}$ 	&COMP    	     	&2.374(1)   & [33]   							\\
 080810  	&Swift   &  $ 13.75\pm 1.10$	& $ 2413_{ -258}^{  +32}$  	&BAND    	    	&3.35(1)    & [34]   							\\
 080804  	&Swift   &  $  6.12\pm 0.56$	& $  997_{  -86}^{  +104}$  	&COMP    	    	&2.2045(1)  & [35]   							\\ 
\hline
\end{tabular}
\tablefoot{
The columns show the name of the GRB, \tninetyrest, \eprest, best fitting model, redshift and redshift references of the 32 GBM GRBs presented here. The keyword ``Swift'' in the mission column denotes whether the GRB triggered both \textit{Swift} and GBM, whereas ``Fermi'' indicates that it was triggered by GBM only.\\
For bursts with * an effective area correction was applied to the BGO with
respect to the NaI detectors. \\
\tablefoottext{a}{The \tninetyrest and \eprest values for GRB~091024 were taken from \citet{gruber11}}
}
\tablebib{
(1)~\citet{tanvir10}	;
(2) \citet{2010GCN.11089....1O}	; 
(3) \citet{2010GCN.10606....1C}	; 
(4) \citet{berger10}	;
(5) \citet{2009GCN.10263....1W}	; 
(6) \citet{2009GCN.10202....1C}	; 
(7) \citet{2009GCN.10065....1C}	;
(8) \citet{2009GCN.10053....1X}	; 
(9) \citet{2009GCN.10031....1C}	; 
(10) \citet{2009GCN..9958....1L}	;
(11) \citet{2009GCN..9947....1F}	;
(12) \citet{2009GCN..9942....1M}	;
(13)	 \citet{2009GCN..9873....1C} ;
(14) \citet{2009GCN..9518....1C}	;
(15) \citet{2009GCN..9409....1T}	;
(16) \citet{2009GCN..9408....1R}	;
(17) \citet{2009GCN..9383....1D}	;
(18) \citet{2009GCN..9382....1R}	;
(19) \citet{2009GCN..9353....1R}	;
(20) \citet{2009GCN..9243....1C}	;
(21) \citet{2009GCN..9250....1W}	;
(22) \citet{tanvir09}	;
(23) \citet{2009GCN..9053....1C} ;
(24) \citet{2009GCN..9028....1C} ;
(25) \citet{2009GCN..8766....1D} ;
(26) \citet{2008GCN..8713....1C} ;
(27) \citet{2008GCN..8542....1B} ;
(28) \citet{2008GCN..8350....1D} ;
(29) \citet{2008GCN..8335....1B} ;
(30) \citet{2008GCN..8304....1C};
(31) \citet{2009AA...498...89G}   ;
(32) \citet{2008GCN..8254....1F} ;
(33) \citet{2008GCN..8191....1V} ;
(34) \citet{2008GCN..8083....1P} ;
(35) \citet{2008GCN..8058....1T} .
}

\end{table*}

\end{document}